\documentclass[sigconf, authorversion=True, screen=True]{acmart}
\usepackage{soul}
\usepackage{url}
\usepackage{booktabs}
\usepackage{algorithm}
\usepackage{algorithmic}
\usepackage{amssymb}
\usepackage{amsmath}
\usepackage{subcaption}
\usepackage{tikz}
\makeatletter
\newcommand{\printfnsymbol}[1]{%
  \textsuperscript{\@fnsymbol{#1}}%
}
\definecolor{db}{RGB}{0, 0, 83}

\newcommand*\circled[1]{\tikz[baseline=(char.base)]{
            \node[shape=circle,fill=db,draw,inner sep=0.5pt] (char) {\textsf{{\color{white}#1}}};}}
            
\setcopyright{acmcopyright}
\copyrightyear{2020}
\acmYear{2018}
\acmDOI{}

\acmConference[KDD Workshop]{KDD Workshop on Machine Learning in Finance}{Aug 24, 2020}{San Diego, CA}
\acmBooktitle{KDD Workshop on Machine Learning in Finance,
  Aug 24, 2020, San Diego, CA}
\acmPrice{15.00}
\setcopyright{rightsretained}

\begin{document}
\title{Multi-stream RNN for Merchant Transaction Prediction}
\author{Zhongfang Zhuang}
\email{zzhuang@visa.com}
\affiliation{
    \institution{Visa Research}
    \streetaddress{385 Sherman Ave}
    \city{Palo Alto}
    \state{California}
    \postcode{94306}
}
\author{Chin-Chia Michael Yeh}
\email{miyeh@visa.com}
\affiliation{
    \institution{Visa Research}
    \streetaddress{385 Sherman Ave}
    \city{Palo Alto}
    \state{California}
    \postcode{94306}
}
\author{Liang Wang}
\email{liawang@visa.com}
\affiliation{
    \institution{Visa Research}
    \streetaddress{385 Sherman Ave}
    \city{Palo Alto}
    \state{California}
    \postcode{94306}
}
\author{Wei Zhang}
\email{wzhan@visa.com}
\affiliation{
    \institution{Visa Research}
    \streetaddress{385 Sherman Ave}
    \city{Palo Alto}
    \state{California}
    \postcode{94306}
}
\author{Junpeng Wang}
\email{junpenwa@visa.com}
\affiliation{
    \institution{Visa Research}
    \streetaddress{385 Sherman Ave}
    \city{Palo Alto}
    \state{California}
    \postcode{94306}
}
\renewcommand{\shortauthors}{Z. Zhuang, et al.}
\begin{abstract}
    Recently, digital payment systems have significantly changed people’s lifestyles. 
    New challenges have surfaced in monitoring and guaranteeing the integrity of payment processing systems.
    One important task is to predict the future transaction statistics of each merchant. 
    These predictions can thus be used to steer other tasks, ranging from fraud detection to recommendation. 
    This problem is challenging as we need to predict not only multivariate time series but also multi-steps into the future. 
    In this work, we propose a multi-stream RNN model for multi-step merchant transaction predictions tailored to these requirements. 
    The proposed multi-stream RNN summarizes transaction data in different granularity and makes predictions for multiple steps in the future.
    Our extensive experimental results have demonstrated that the proposed model is capable of outperforming existing state-of-the-art methods.
    
\noindent \textbf{Topic Area:} Application - Monitoring, Forecasting 
\end{abstract}

\maketitle

\section{Introduction}
\label{sec-introduction}
Advanced digital payment technology has enabled billions of transactions to be processed every second. 
One challenge these systems face is detecting the irregular transaction behaviors of merchants that deviate from the historical data. 
While many of these deviations may be harmless, some may signal serious underlying issues, ranging from connection issues between point-of-sale (POS) and the payment processing centers to money laundering. 
Thus, detecting deviated behaviors is not only an important task for both merchants and payment processing companies but also a responsibility for every participant of the payment ecosystem.

To detect deviated behaviors, one crucial step is to study the transaction data of every merchant from the past and estimate the future. 
\begin{figure}[t]
    \centering
    \includegraphics[width=\linewidth]{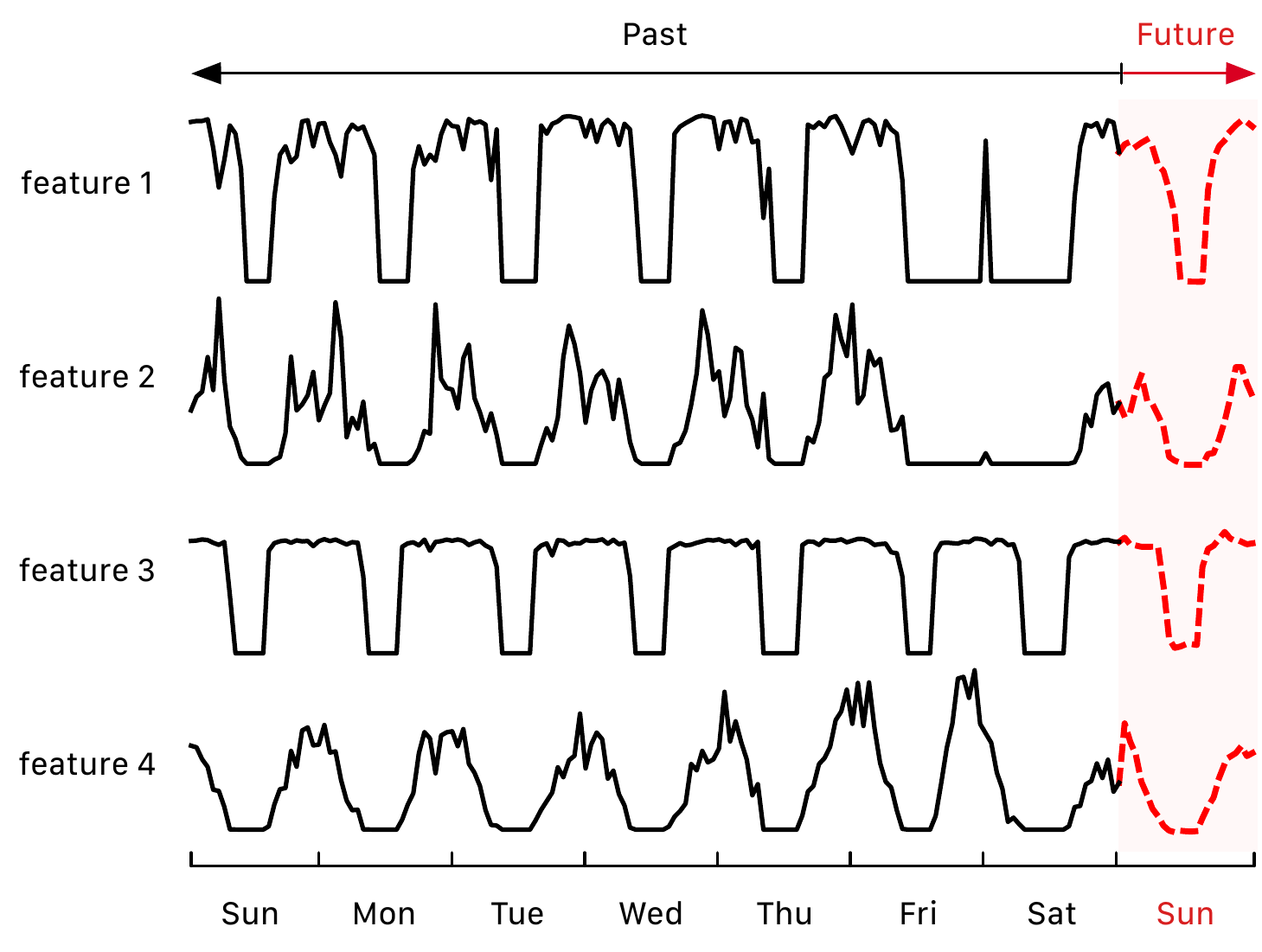}
    \caption{An example of time series prediction in the context of the merchant transaction history. Four features exhibit \textit{daily} recurring behaviors.}
    \label{figm-transact-history}
    \vspace{-5mm}
\end{figure}
Transaction data in real-world applications are often stored as multivariate time series where each dimension is a time-varying feature. These features, such as {the hourly transaction amount}, {the hourly number of transactions} and {hourly averaged transaction amount}, are extracted and aggregated for analysis. 
We illustrate one merchant's transaction history in a week in  Figure~\ref{figm-transact-history}. 
\begin{figure*}[t]
    \centering
    \includegraphics[width=0.75\textwidth]{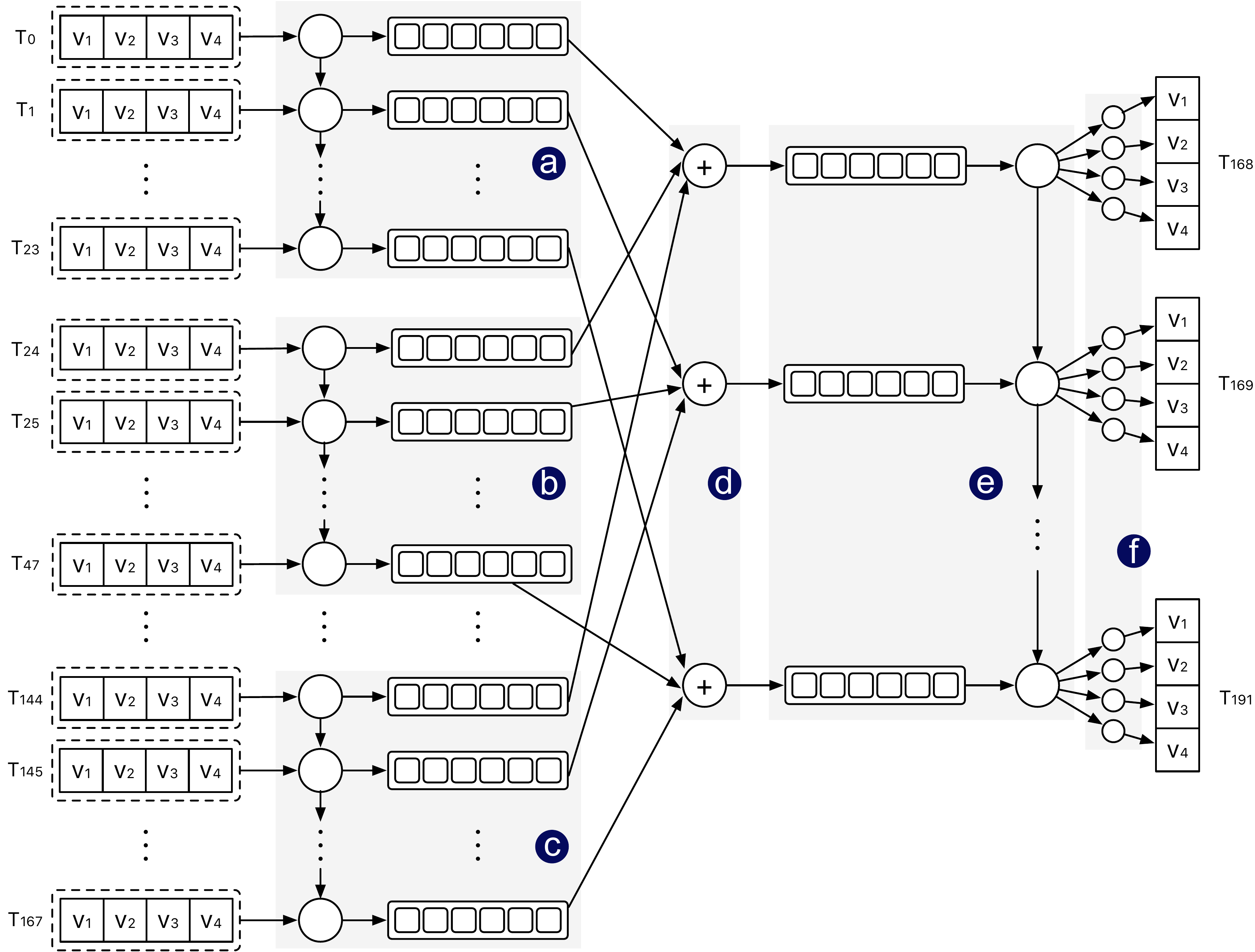}
    \caption{An overview of our proposed model. Components are explained in Section~\ref{sec-proposed-model}.}
    \label{figm-model-overview}
\end{figure*}

Time series prediction is a well-studied problem~\cite{box1970time}. However, predicting multiple time series features in real-world scenarios remains challenging as the real-world data is dynamic and may be impacted by various unpredictable factors, such as traffic and weather. Recent works on time series prediction~\cite{taieb2015bias,wen2017multi} take a new look at this problem from the neural network perspective. Specifically, Taieb and Atiya~\cite{taieb2015bias} proposes to train a neural network with a target consisting of multi-steps into the future; Wen \textit{et al.}~\cite{wen2017multi} approaches this problem from sequence-to-sequence perspective. 
However, in these methods~\cite{taieb2015bias,wen2017multi}, each feature is treated independently without assuming any dependencies among them which contradicts our domain knowledge in transaction data as different features in transaction data are inter-dependent. 

In this work, we propose to use a multi-stream RNN model to tackle the merchant transaction prediction task, where each stream is responsible for the data in each day of the week based on observed merchant behaviors. Our approach first aggregates transaction data at a fine granularity (e.g., day) and then augment it with another RNN to capture the pattern at a higher level. 
Different from the existing work in computer vision research~\cite{zhang2019deep}, where multi-stream RNNs are used to predict each time step for action labels, the goal of using multi-stream RNN in this work is to summarize and preserve as much transaction information as possible for future long-term predictions. 

We summarize the contributions of this work as follows: 
\begin{itemize}
    \item We propose and analyze the problem of multivariate multi-step merchant transaction prediction. 
    \item We design a multi-stream RNN model to tackle the multi-step prediction task. 
    \item We demonstrate the effectiveness of our proposed model by comparing it with various baseline methods on real-world aggregated transaction data. 
\end{itemize}
\section{Multi-stream RNN}
\label{sec-proposed-model}
In this section, we first use Figure~\ref{figm-model-overview} to explain the architecture of the proposed model in Section~\ref{sec-msrnn}.
Next, we discuss design variations for some of the model components in Section~\ref{sec-variations}.

\subsection{Architecture}
\label{sec-msrnn}
The input to our model is the hourly aggregated transaction data of 168 hours ($T_{0} \sim T_{167}$) in a form of multivariate time series, and the output of our model is the predicted multivariate time series for the next 24 hours ($T_{168} \sim T_{191}$).
The core components of multi-stream RNN are:
\begin{enumerate}
    \item Daily RNNs. Each daily RNN is tasked to process each day's transaction data (e.g., \circled{a}-\circled{c}). 
    \item Merge layers (i.e., \circled{d}). Each merge layer is responsible for aggregating the information from the hourly output of each daily RNNs. 
    The hourly outputs associated with the same hour of the day are aggregated together. 
    \item Weekly RNN (i.e., \circled{e}) connects to the previous 24 merge layers to capture the high-level patterns in the input time series.
    \item The fully-connected layers (i.e., \circled{f}) that make predictions for each of the 24 temporal steps individually. 
\end{enumerate} 

One main component in our model is the daily RNN, with the goal of summarizing the daily merchant time series patterns. 
In particular, we use either GRU or LSTM as the daily RNN. 
We denote each daily RNN here as: 
\begin{equation}
    \mathbf{h}_i^{(t)} = \text{\textsf{RNN}}_i(\mathbf{x}_t)
\end{equation}
where $\mathbf{x}_t$ denotes the values of merchant features at time $t$, $\text{\textsf{RNN}}_i$ is the $i$-th RNN in the multi-stream schema and $\mathbf{h}_i^{(t)}$ is the latent representation of the time series of the $i$-th day of the week (at time $t$). 
For example, $\text{\textsf{RNN}}_1$ is only responsible of processing $T_0 \sim T_{23}$ transaction data.

Each RNN used here has the same number of hidden dimensions. 
A dropout layer is augmented after each RNN with a fixed 0.2 dropout rate. 

Next, the hidden states $\mathbf{h}_i^{(t)}$ are aggregated by merge layers \circled{d} where each merge layer captures the patterns across every day of the week for different hours: 
\begin{equation}
    \mathbf{s}_t = \mathbf{h}_0^{(t)} + \mathbf{h}_1^{(t)} + ... + \mathbf{h}_6^{(t)}
\end{equation}
where $t \in [0, 23]$ and $\mathbf{h}_0^{(t)} \sim \mathbf{h}_6^{(t)}$ contains the information associated with a given hour $t$ for each day in the week. 

On a higher level, we utilize \circled{e} to learn the patterns across output of each merge layer. \circled{e} can be written as: 
\begin{equation}
    \mathbf{h}_\text{s}^{(t)} = \text{\textsf{RNN}}_\text{s}(\mathbf{s}_t)
\end{equation}

Lastly, we utilize $\mathbf{h}_\text{s}^{(t)}$ to predict the future values of each feature by appending fully connected layers after $\text{\textsf{RNN}}_\text{s}$ (i.e., \circled{f}). 
Instead of predicting values of all features together, we utilize separated linear fully connected layers to predict each of the $k$-th feature individually. These fully connected layers are: 
\begin{equation}
    \mathbf{v}_k^{(t)} =  \mathbf{W}_k\mathbf{h}_\text{s}^{(t)} + \mathbf{b}_k
\end{equation}
where $\mathbf{W}_k$ is the weight matrix and $\mathbf{b}_k$ is the bias term of fully connected layer. 
\subsection{Model Variations} 
\label{sec-variations}
\begin{figure}[t]
    \centering
    \includegraphics[width=0.8\linewidth]{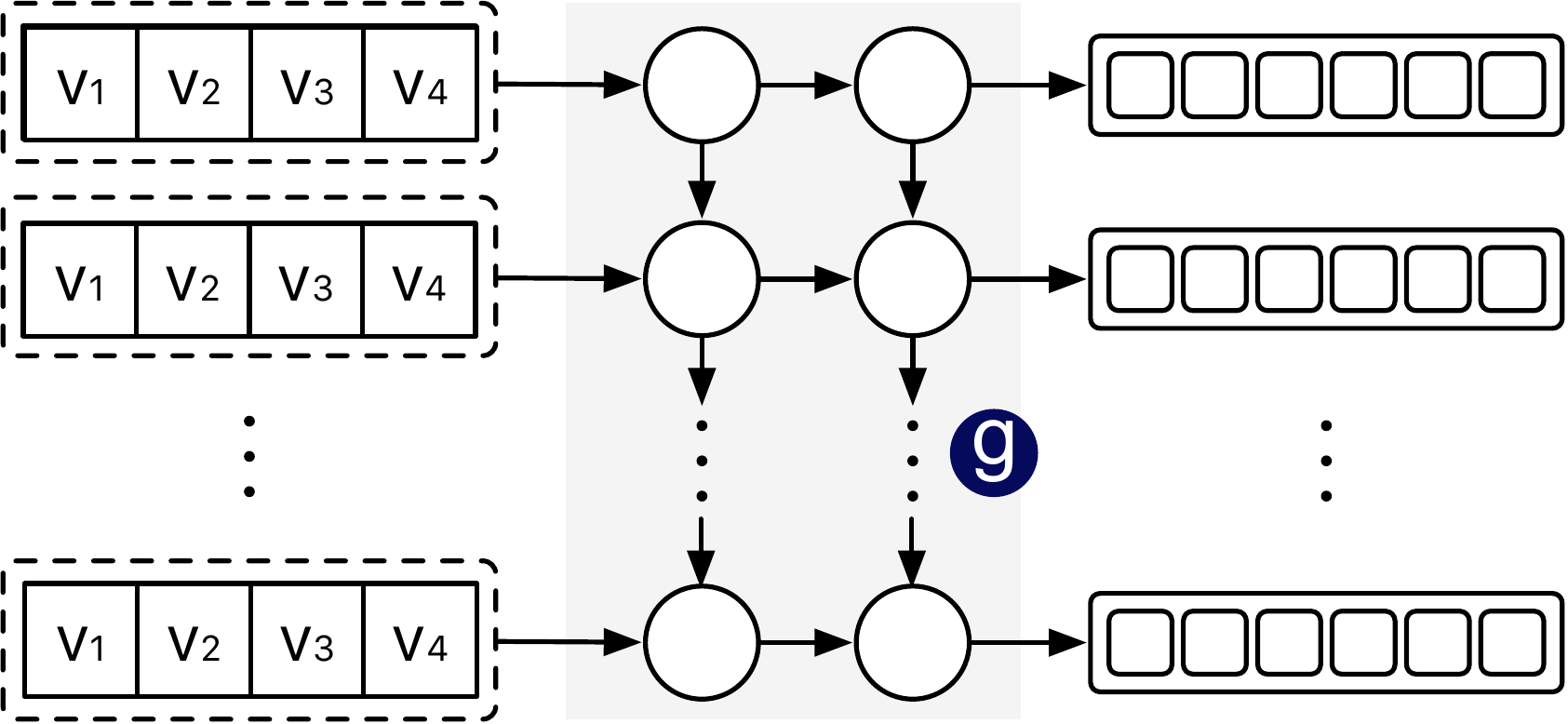}
    \caption{Stacked RNN for complex time series input. }
    \label{figm-stackedrnn}
\end{figure}


\noindent \textbf{Stacked RNN. }
In the original design, we only use one RNN layer as depicted in Figure~\ref{figm-model-overview}. However, time series that contains more complicated patterns requires more complicated RNN structure, such as stacked RNN, to capture the patterns. Thus, one variation of our model is to replace the daily RNN layers (e.g., \circled{a} $\sim$ \circled{c}) with the new stacked RNN layers (\circled{g}) in Figure~\ref{figm-stackedrnn}.

\begin{figure}[t]
    \centering
    \includegraphics[width=0.6\linewidth]{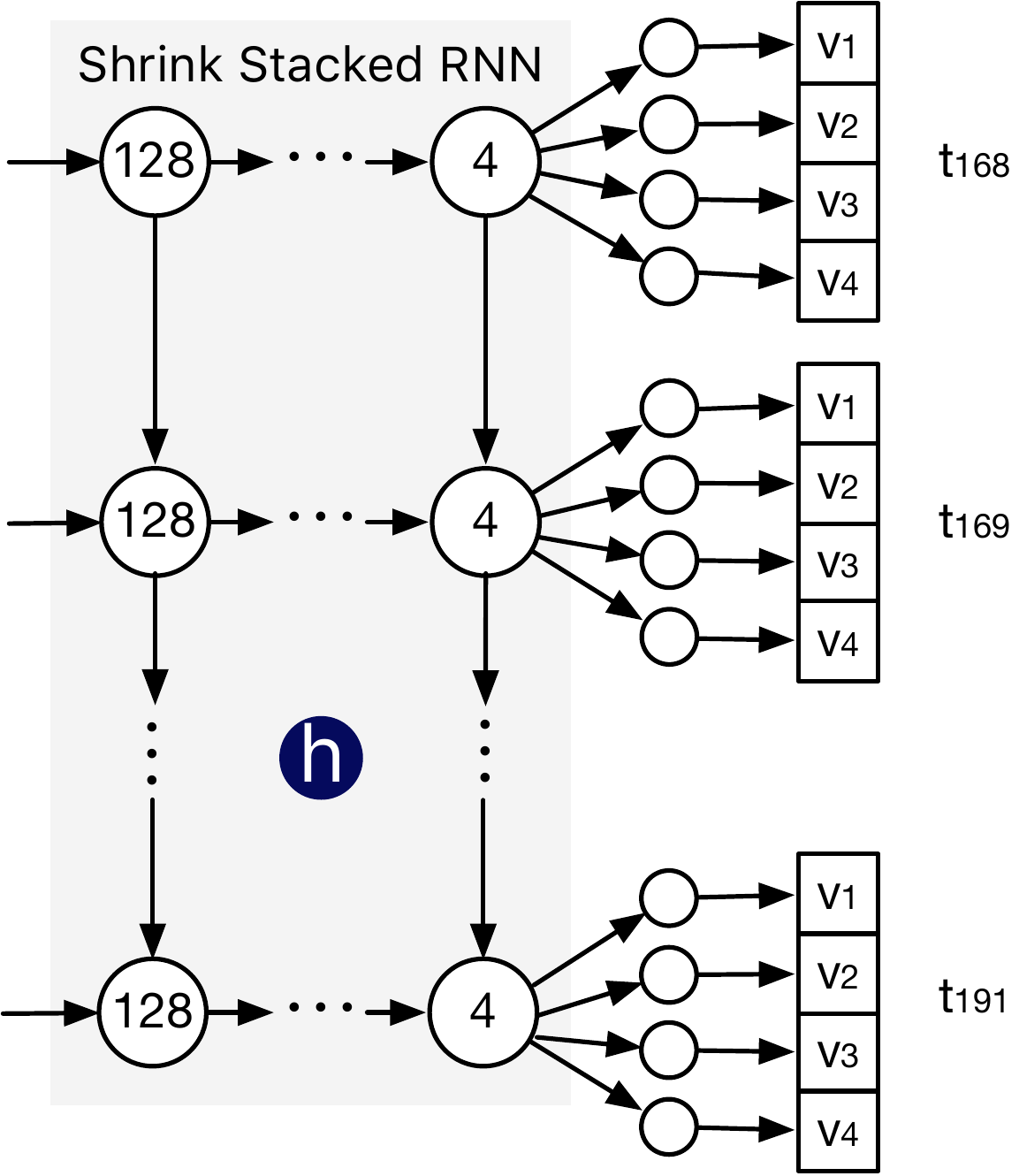}
    \caption{Shrink stacked RNN for output. }
    \label{figm-shrinkrnn}
\end{figure}

\noindent \textbf{Shrink Stacked RNN for Output. }
Another variation of our model is in the higher level weekly RNN layer (i.e., \circled{e}). Different from the vanilla design of the weekly RNN layer, we now have multiple RNN layers to \textit{gradually} reduce the dimensionality of the RNN layers (denoted as \circled{h} in Figure~\ref{figm-shrinkrnn}). We shrink the dimensions of each output layer using the following rule: 
\begin{equation}
\label{eq-shrinkrnn}
    d_{i+1} = 
    \begin{cases}
    \lfloor d_{i} / r \rfloor, &\text{otherwise} \\ 
    m, &\text{if} \lfloor d_{i} / r \rfloor = 0 
    \end{cases}
\end{equation}
where $d_{i+1}$ and $d_{i}$ are the dimensions of the $(i+1)$-th and $i$-th layers, respectively. $r$ represents the rate of shrinkage. $m$ is the minimal dimension of this stack when $\lfloor d_{i} / r \rfloor = 0$. This is to ensure the last output RNN layer has a valid number of dimensions.

\section{Experiments}
\subsection{Description of the Datasets}
\label{sec-exp-dataset}
We have organized four different datasets where each dataset consists of merchants from one of the following categories: department store, restaurants, sports facility, and medical services, denoted as C1 $\sim$ C4, respectively.

For each category, we randomly select 2,000 merchants located within California, United States.
The time series datasets consist of four features, each is produced by computing the hourly aggregation of the following statistics: number of approved transactions, number of unique cards, a sum of the transaction amount, and rate of the approved transaction.
The training data consists of time series data during November 1--23, 2018 and the test data is time series during November 24--30, 2018.

As mentioned in Section~\ref{sec-introduction}, the goal of the system is to predict the next 24 hours given the last 168 hours (i.e., seven days).
We predict every 24 hours in the test data by supplying the latest 168 hours to the system.
For example, the transaction data of 168 hours in \texttt{Week-10} is used to predict the values of 24 hours on the Monday of \texttt{Week-11}.

\subsection{Training Details}
\noindent \textbf{Model Parameters.} 
In the recurrent layers, we use 256 as the number of dimensions. Based on our experiments, we choose \texttt{sigmoid} as the activation function of the input RNN layers and \texttt{relu} as the activation function of the output RNN layers. We use \texttt{rmsprop} with learning rate of 0.001 as the optimizer. Batch size is set to 256 as a balance between computing resource utilization and model quality. We chose $r=2$ and $m=4$ in Equation~\ref{eq-shrinkrnn} for the experiments using Shrink RNN output layers. The maximum number of training epochs is 100 and early stopping with a patience of 5 is used to avoid overfitting. 
\begin{table*}[t]
\centering
\caption{Experiment result summary. }
\label{tab-exp-result}
\resizebox{\textwidth}{!}{
\large
\begin{tabular}{lcccccccccc} 
\toprule
          &\multicolumn{5}{c}{RMSE} &\multicolumn{5}{c}{Normalized RMSE}\\ 
\cmidrule(lr){2-6} \cmidrule(lr){7-11}
          & \begin{tabular}[c]{@{}c@{}}Department \\Store\end{tabular} & Restaurants & Sports & Medical & Average & \begin{tabular}[c]{@{}c@{}}Department \\Store\end{tabular} & Restaurants & Sports & Medical & Average  
          \\
\midrule
Linear Model & 3.7006 & 3.7096 & 4.3325 & 3.7536 & 3.8741 & 10.5704 & 5.5397 & 10.2111 & 9.9309 & 9.0630 \\
Linear Model (L2-regularized) & 3.6999 & 3.7094 & 4.3324 & 3.7532 & 3.8737 & 10.5669 & 5.5393 & 10.2107 & 9.9300 & 9.0617 \\
Nearest Neighbor              & 5.9784 & 5.7083 & 8.6402 & 5.4327 & 6.4399 & 11.0063 & 6.8305 & 11.2740 & 9.5840 & 9.6737 \\
Random Forest & 3.6918 & 4.2933 & 4.8370 & 3.9903 & 4.2031 & \textbf{8.9944} & 6.0096 & 10.5885 & 9.2405 & 8.7082 \\
RNN (1-layer LSTM) & 3.6920 & 3.8042 & 4.3263 & \textbf{3.4097} & 3.8931 & 10.6102 & 5.6537 & 10.1720 & 9.1584 & 9.2342 \\
RNN (2-layer LSTM) & 3.8138 & 3.7908 & 4.3121 & 3.3720 & 3.9190 & 9.9693 & 5.7273 & 10.2029 & 8.9879 & 9.0963 \\
RNN (1-layer GRU) & 3.5531 & \textbf{3.6632} & 4.3158 & 3.5113 & 3.8726 & 10.2535 & 5.5724 & 10.1210 & 9.6771 & 9.2512 \\
RNN (2-layer GRU) & 3.8311 & 3.7381 & 4.3193 & 3.4631 & 3.9398 & 10.5512 & 5.6134 & 10.2579 & 9.1511 & 9.2519 \\
\midrule
MS-RNN (1-layer LSTM) & 3.5647 & 3.7825 & 4.2409 & 3.8037 & 3.8479 & 10.1082 & 5.4290 & 10.4083 & 8.5998 & 8.6363 \\
MS-RNN (1-layer GRU) & \textbf{3.5521} & 3.7050 & \textbf{4.1914} & 3.5186 & \textbf{3.7418} & 11.1979 & \textbf{5.3735} & \textbf{9.5052} & \textbf{8.3351} & \textbf{8.6029} \\ 
\bottomrule
\end{tabular}}
\end{table*}


\noindent \textbf{Training Goal.} 
The training goal is to minimize the mean square error between the predicted and true values of the features $\mathbf{v}_k^{(t)}$ of the next 24 hours within a mini-batch. The training goal can be written as: 
\begin{equation}
    \frac{1}{n}\sum_{i=1}^{n}\sum_{k=1}^{4}\sum_{t=0}^{23}\left(\mathbf{v}_k^{(t)} - \widehat{\mathbf{v}_k^{(t)}} \right)
\end{equation}
where $k=1, \cdots, 4$ represents the four features, $t=0, \cdots, 23$ represents the 24 hour period.  

\subsection{Compared Methods}

\begin{figure}[t]
    \centering
    \includegraphics[ width=0.475\textwidth]{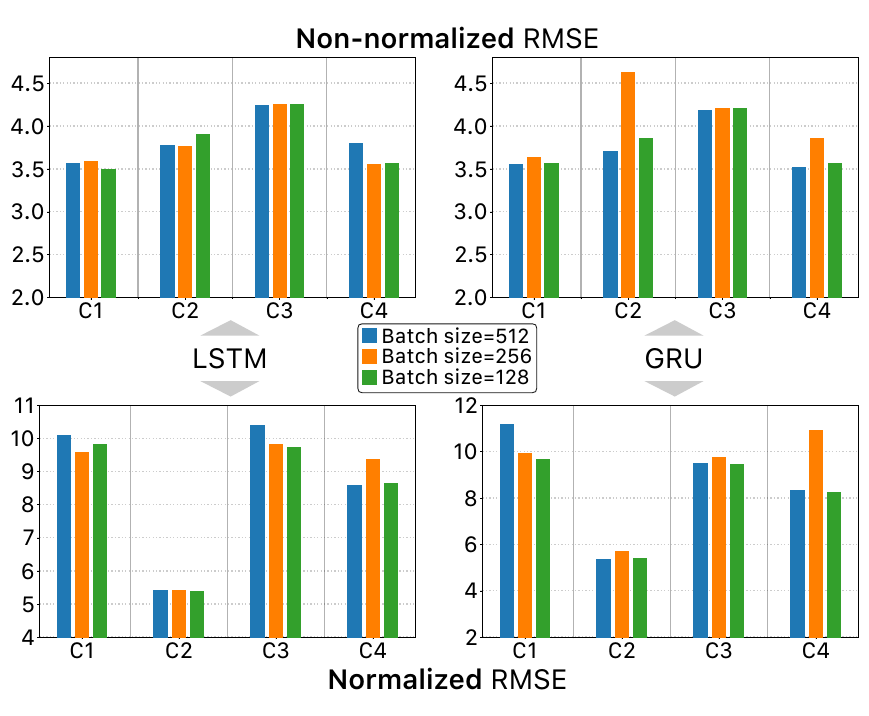}
    \caption{\textbf{Non-normalized} and \textbf{Normalized} RMSE results with various batch sizes.}
    \label{fige-batch-size}
\end{figure}

\begin{itemize}
  \item \textbf{Linear Model} is one of the simpler models for solving regression problems.
  It predicts a value by linearly combining the input vector.
  Since we are predicting the four features for the next 24 hours (i.e., 96 values total), we train 96 linear models and each model predicts one value.
  The input to each of the models is a 672-sized vector consisting of the four time series from the last 168 hours.
  We also test linear models under both l2-regularized and non-regularized settings.
  When l2 regularization is used, the parameter associated with the strength of regularization is found using three-fold cross-validation.
  \item \textbf{Nearest Neighbor} is another simple method for solving regression problems.
  It predicts the future by finding the nearest neighbor of the current time series from the past.
  Once the nearest neighbor is located, the next 24 hours of the nearest neighbor can be used as the prediction of the current time series.
  This method is one of the more prevalent methods in time series data mining~\cite{rakthanmanon2012searching}.
  \item \textbf{Random Forest} is an ensemble method utilizing both bootstrap aggregating and random subspace method~\cite{ho1995random,breiman2001random} for training a set of decision trees.
  The input to the model is once again a 672-sized vector consisting of the four time series from the last 168 hours, and the output of the model is a 96-sized vector consisting of the four time series for the next 24 hours.
  The hyper-parameters associated with the random forest is found based on the estimated error of three-fold cross-validation.
  \item \textbf{Recurrent Neural Network} is a popular artificial neural network model for modeling sequence data.
  We use one or two layers of RNN to encode the input time series, then we use a Multi-layer Perceptron (MLP) to predict the time series for the next 24 hours following the state-of-the-art~\cite{taieb2015bias,wen2017multi}.
  Similar to the proposed method, we test the model with both LSTM~\cite{gers1999learning} and GRU~\cite{cho2014learning} recurrent architecture.
\end{itemize}

\begin{figure}[t]
    \centering
    \includegraphics[ width=0.475\textwidth]{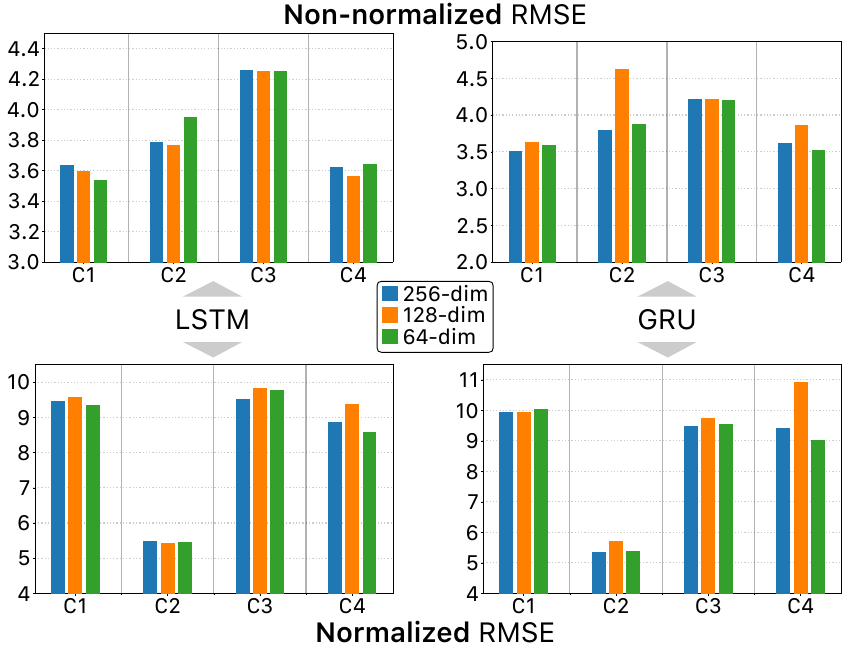}
    \vspace{-3mm}
    \caption{\textbf{Non-normalized} and \textbf{Normalized} RMSE results with various hidden dimensions.}
    \label{fige-hidden-dimension}
\end{figure}

\subsection{Results}
\label{sec-exp-results}

We have evaluated the proposed method and the baseline methods on the four datasets described in Section~\ref{sec-exp-dataset}. 
In addition to the more conventional root mean square error (RMSE), we also measure the performance in normalized RMSE.
We compute normalized RMSE by $z$-normalizing~\cite{rakthanmanon2012searching} each dimension of the predicted time series before computing the RMSE.
Such performance measurement focus on how much the shape of the predicted time series is different from the ground truth time series.
The experiment result is summarised in Table~\ref{tab-exp-result}.
For the proposed method, we have tested the performance with either 1-layer LSTM or 1-layer GRU.

When we consider only the baseline methods, a simple L2-regularized linear model outperforms all the other \textit{non}-deep learning methods and is comparable to deep learning-based methods in averaged RMSE.
For averaged normalized RMSE, random forest outperforms all other baseline methods, even the state-of-the-art RNN-based solution which shows how normalized RMSE could provide us an alternative view on the predictions provided by different methods.
Nevertheless, both the averaged RMSE and averaged normalized RMSE suggest that MS-RNN with either LSTM or GRU outperforms all baseline methods.

MS-RNN outperforms the RNN on two out of four merchant types when we look closely at the result of each merchant type for RMSE, while MS-RNN outperforms the RNN on three out of four merchant types for normalized RMSE. 
Coupled with the fact that the improvement in RMSE is 3.4\% while the improvement in normalized RMSE is 5.4\%, MS-RNN is more capable of modeling the shape of the time series compared to the state-of-the-art RNN method.
Such improvement can be attributed to the model architecture as the multi-stream design helps the model attending more to the details of the input time series.

\subsection{Parameter Study}
We test one parameter associated with the model architecture and training process in each set of experiments, including batch size, number of hidden dimensions, number of RNN layers and shrink stacked RNN.

\noindent \textbf{Batch Size. }
Three different settings of batch size (i.e., 128, 256 and 512) are tested for the proposed MS-RNN with either LSTM or GRU.
Similar to Section~\ref{sec-exp-results}, we report both the RMSE and normalized RMSE.

The result is shown in Figure~\ref{fige-batch-size}.
There is no apparent trend when considering the performance of the model with respect to the batch size.
In other words, the proposed architecture is not sensitive to the parameterization of the optimization process.

\noindent \textbf{Number of Hidden Dimensions. }
We have considered three different settings for the number of hidden dimensions (i.e., 64, 128 and 256).
The experiment result is summarised in Figure~\ref{fige-hidden-dimension}.
There is no clear trend when considering the performance with respect to the number of hidden dimensions which suggested that setting the number of hidden dimensions to 64 is sufficient for these datasets. 
\begin{figure}[t]
    \centering
    \includegraphics[width=0.495\textwidth]{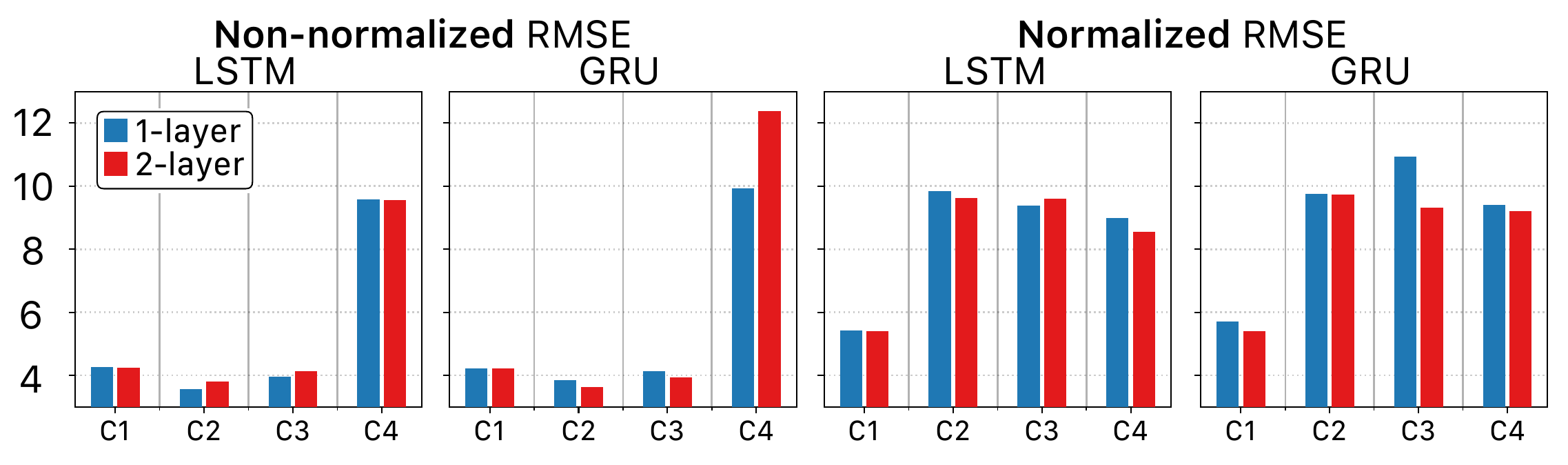}
    \vspace{-8mm}
    \caption{Effect of number of layers (shared $y$-axis). }
    \label{fige-num-layers}
\end{figure}
\begin{figure}
    \centering
    \includegraphics[width=0.49\textwidth]{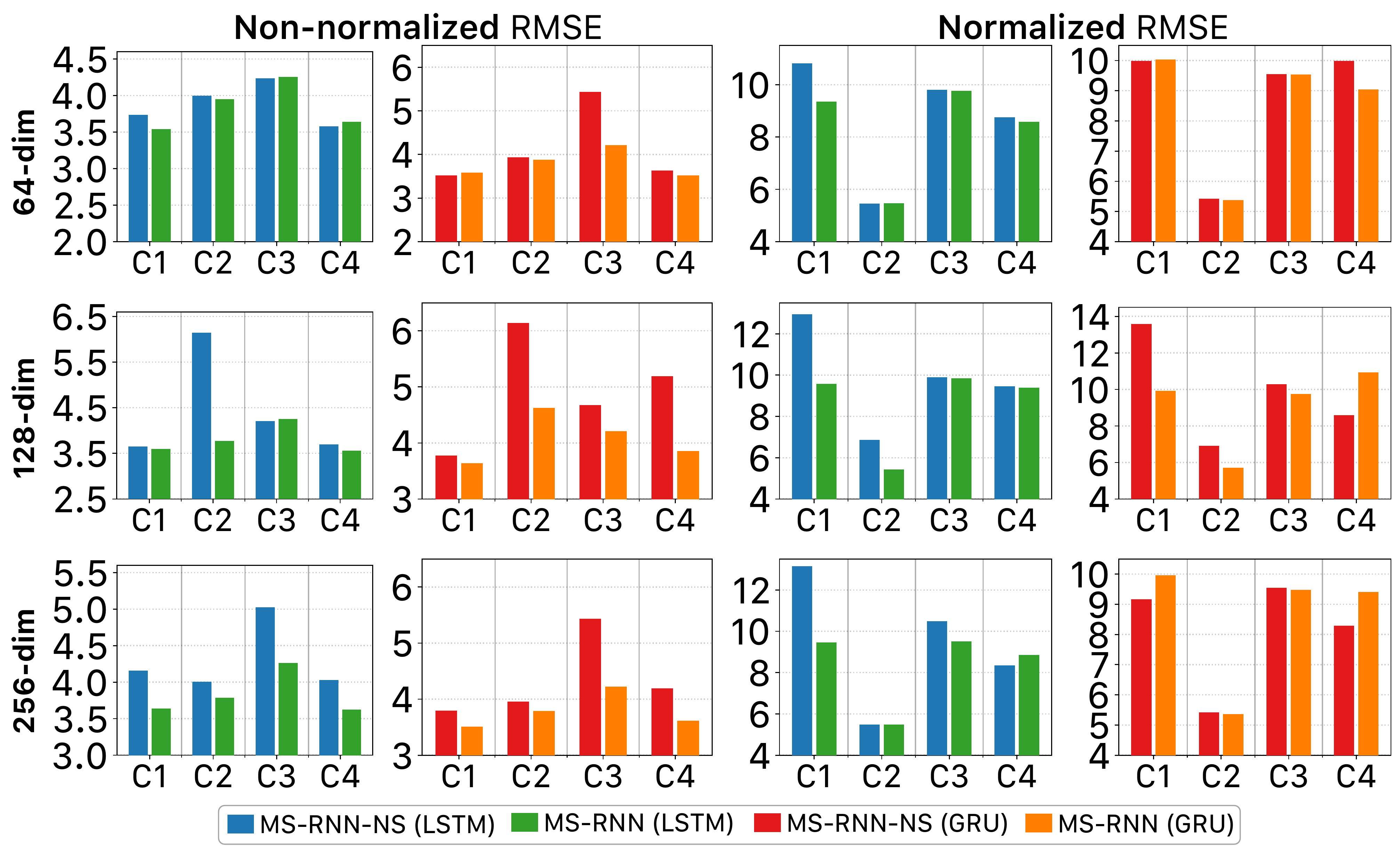}
    \caption{The effectiveness of shrink stacked RNN with various parameter settings. }
    \label{fige-ssrnn}
\end{figure}

\noindent \textbf{Number of RNN layers. }
We compare the performance difference between the 1-layer variant and 2-layer variant for MS-RNN with both LSTM and GRU; the result is presented in Figure~\ref{fige-num-layers}.
The increment in the number of layers helps our model to model the shape better (higher normalized RMSE); however, it also slightly degraded the RMSE on all datasets for LSTM and greatly degraded the RMSE on medical merchants for GRU.
The choice for this parameter would depend on whether RMSE is more important than normalized RMSE or not.
\begin{figure*}[t]
    \centering
    \includegraphics[width=\linewidth]{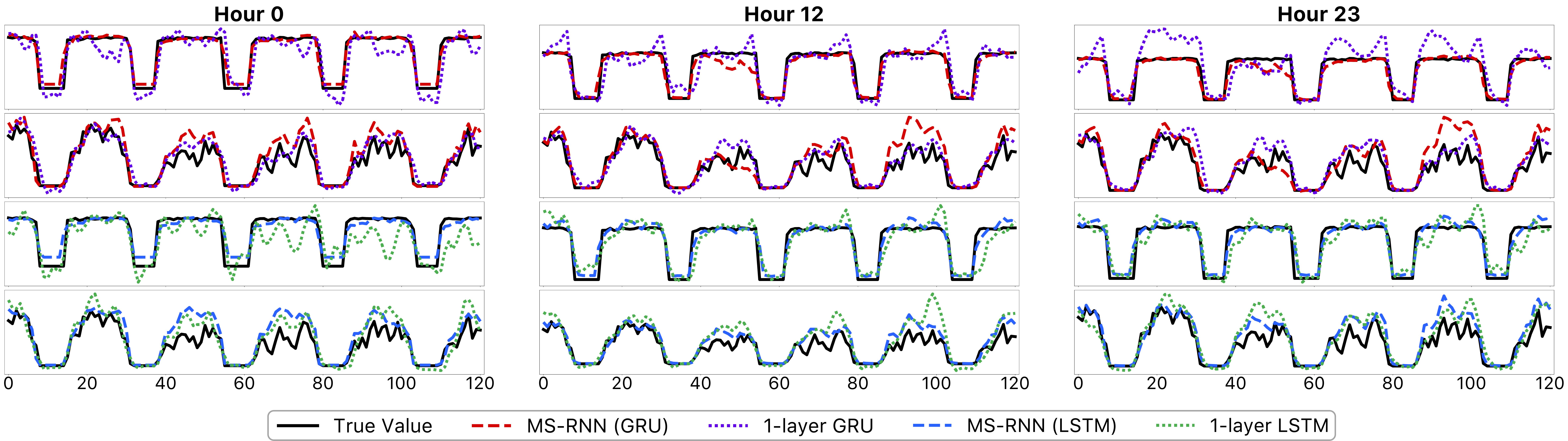}
    \caption{The true and predicted time series of a randomly chosen merchant for the immediate future (i.e., hour 0), near future (i.e., hour 12) and far future (i.e., hour 23). All values are normalized to $[0, 1]$ range.}
    \label{fige-tspred}
\end{figure*}
\begin{figure}[t]
    \centering
    \includegraphics[width=0.45\textwidth]{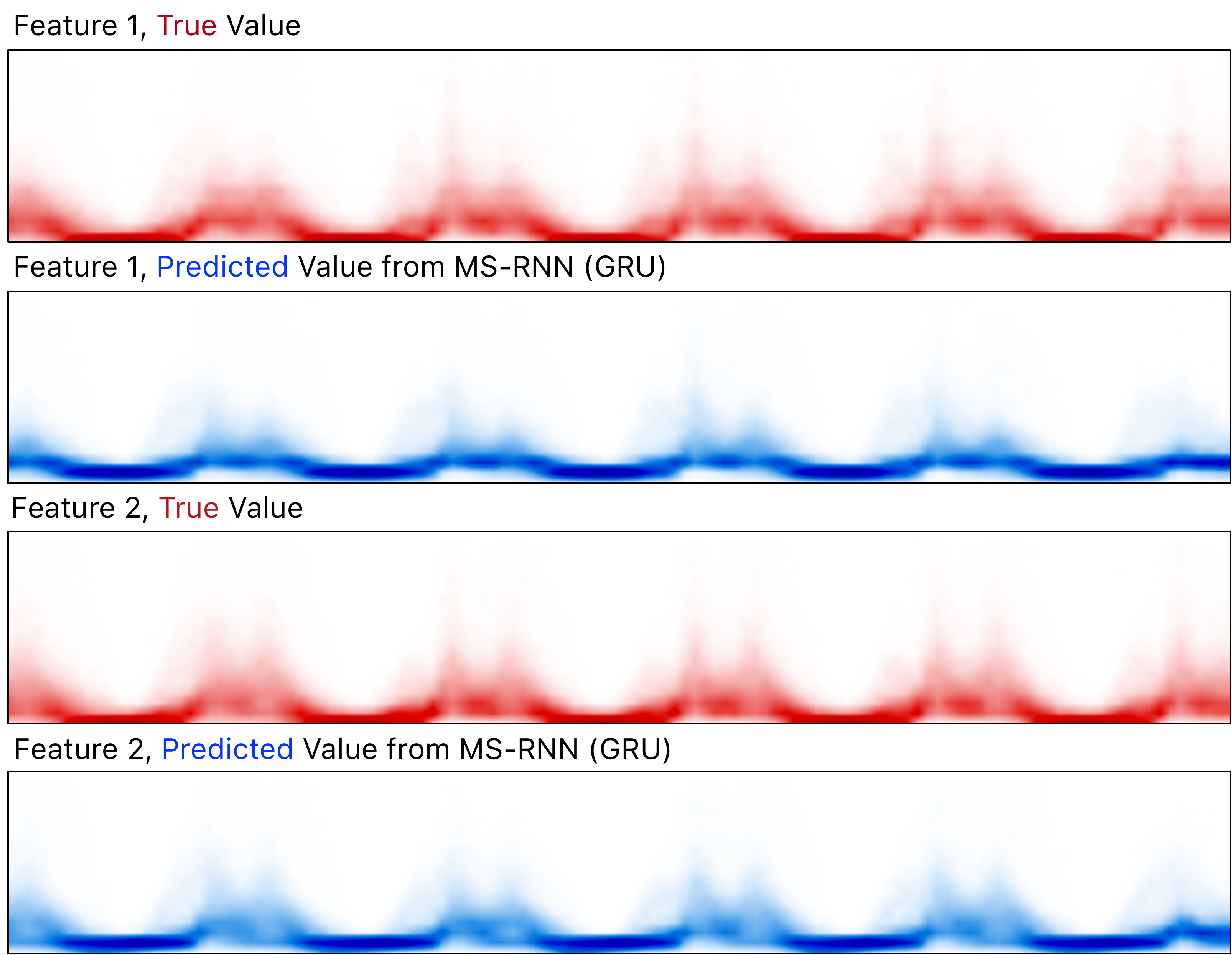}
    \caption{A density plot of the \textbf{last} predicting step ($t+23$) of all merchants in the restaurant category. All values are normalized to $[0, 1]$ range. }
    \label{fige-density}
\end{figure}

\noindent \textbf{Shrink Stacked RNN. }
We tested the effectiveness of shrink stacked RNN on 6 different architecture settings (LSTM or GRU with 3 different numbers of hidden dimensions) on 4 different datasets.
In other words, we compare the MS-RNN with or without shrink stacked RNN under 24 different settings.
The comparison under the 24 experiment settings is organized in Figure~\ref{fige-ssrnn}.
The effect is positive for adding shrink stacked RNN to the MS-RNN.
When we look closely at the non-normalized RMSE results, MS-RNN with shrink stacked RNN outperforms their counterparts without shrink stacked RNN in 19 out of 24 setups.
Additionally, when the normalized RMSE results are considered, MS-RNN with shrink stacked RNN achieves better performance compared to their counterparts in 17 out of 24 setups.
The shrink stacked RNN design \textit{does} help the MS-RNN converge to a better solution.

\subsection{Case Studies}
In addition to the quantitative analysis of the MS-RNN model, we perform a qualitative analysis of the time series prediction output by the model.
Particularly, we plot the predicted time series of MS-RNN along with both of the ground truth and the state-of-the-art RNN methods; the specific variant that we plot for both of these methods is the 1-layer GRU and 1-layer LSTM.

In Figure~\ref{fige-tspred}, we plot the predicted time series of a randomly chosen merchant for the immediate future (i.e., hour 0), near future (i.e., hour 12) and far future (i.e., hour 23).

We demonstrate the associated time series for the GRU variants of each model in the top two rows and LSTM variations of each model in the bottom two rows of Figure~\ref{fige-tspred}. 
Each row within the respective group corresponds to a different feature. 
We can see that the MS-RNN with either GRU or LSTM is better than the state-of-the-art RNN in all three time points and all two features especially for the hour 23 in the first feature.
When we focus on the latter two rows, which shows the prediction results for both MS-RNN and RNN's LSTM variants, the conclusion is similar that MS-RNN produces a prediction of higher quality comparing to RNN.
Such a visual examination confirms that the prediction from the MS-RNN model is better than the state-of-the-art RNN model.

To provide a dataset-wise holistic visualization of the prediction using MS-RNN, we present the density plot for our model against the ground truth in Figure~\ref{fige-density}. 

We discretize the plot region into grid cells and record the number of points falling into each cell when overlaying the hour 23 prediction of all restaurant merchants.
The color from white to red/blue reflects the increasing count from 0 to the maximum samples for either the ground truth or MS-RNN.
It can be seen that the density plot for MS-RNN matches with that of the ground truth, which further assures the prediction capability of MS-RNN. 

\section{Conclusion and Future Work}
In this work, we identified a real-world problem of predicting multiple features in multiple future consecutive time steps. We proposed a novel and flexible multi-stream RNN based solution to capture merchant transaction patterns. Our extensive experiments have demonstrated that the proposed framework outperforms existing baseline methods under various configurations. 

Predicting multiple steps in the future on multivariate time series is an interesting topic in the FinTech industry. One limitation of this work is that we use only one week's data to predict the following day's transaction. A possible future elaboration of this work could be to modify the network and utilize transaction data from a longer period (e.g., a month). 

Another interesting research direction to tackle this problem would be to explore non-recurrent neural networks, such as convolutional neural networks (CNN). CNN is potentially useful for this problem for two reasons: (1) CNN has a simpler structure compared to RNN, which simplifies the training process. (2) Intuitively, CNN may also fit well with the sliding window problem setting in time series research.

\bibliographystyle{ACMRefStyle}
\newpage
\bibliography{contents/ref.bib}
\end{document}